\documentclass[aps,prl,twocolumn,groupedaddress,draft,showpacs,intlimits,amsmath,amssymb,floats]{revtex4}
\usepackage{bm}
\usepackage[final]{graphicx}
\usepackage{epsfig}
\begin{document}
\title{Anisotropic electron-phonon interaction in the cuprates}
\date{\today}
\author{T. P. Devereaux}
\email{tpd@lorax.uwaterloo.ca}
\homepage{http://www.sciborg.uwaterloo.ca/~tpd/}
\affiliation{Department of Physics, University of Waterloo,
Canada, Ontario N2L 3GI}
\author{T. Cuk}
\email{tanjacuk@stanford.edu}
\affiliation{Dept. of Physics, Applied Physics and Stanford
Synchrotron Radiation Laboratory, Stanford University, California
94305, USA}
\author{Z.-X. Shen}
\email{zxshen@stanford.edu} \homepage{http://arpes.stanford.edu/}
\affiliation{Dept. of Physics, Applied Physics and Stanford
Synchrotron Radiation Laboratory, Stanford University, California
94305, USA}
\author{N. Nagaosa}
\email{nagaosa@appi.t.u-tokyo.ac.jp}
\homepage{http://appi.t.u-tokyo.ac.jp/index-e.htm}
\affiliation{CREST, Department of Applied Physics, University of Tokyo,
Bunkyo-ku, Tokyo 113-8656, Japan}

\begin{abstract}
We explore manifestations of electron-phonon coupling on the electron
spectral function for two phonon modes in the
cuprates exhibiting strong renormalizations with temperature and
doping. Applying simple symmetry considerations and kinematic
constraints, we find that the out-of-plane, out-of-phase O
buckling mode ($B_{1g}$) involves small momentum transfers and
couples strongly to electronic states near the anti-node
while the in-plane Cu-O breathing modes involve large
momentum transfers and couples strongly to nodal electronic
states. Band renormalization effects are found to be strongest in the
superconducting state near the anti-node, in full
agreement with angle-resolved photoemission spectroscopy (ARPES) data.
\end{abstract}


\maketitle

The electron-phonon interaction in metals is generally considered
to be isotropic. Recent developments in the cuprates however
raises important questions about this assumption. The renewed
interest stems from the observation of bosonic band
renormalization effects in a wide range of p-type
materials\cite{Review}. This effect is found to be dramatically
enhanced below T$_{c}$ near the $(\pi,0)$ regions in the Brillouin
zone (BZ)\cite{Anti-Nodal}.
Such a momentum and temperature dependence appears to support the
idea that the bosonic renormalizations in p-type materials is due
to a coupling of the electrons to the 41 meV spin resonance mode
that appears below T$_{c}$\cite{neutron,pi_coupling}.

On the other hand, it has been argued that the bosonic
renormalization effect may be reinterpreted as a result of a
coupling to the half-breathing in-plane Cu-O bond-stretching
phonon for nodal directions\cite{Nodal} and the out-of-plane
out-of-phase O buckling $B_{1g}$ phonon for anti-nodal
directions\cite{Tanja}. This re-interpretation has the clear
advantage over the spin resonance in that it naturally explains
why the band renormalization effect is observed under three
circumstances: 1) in materials where no spin mode has been
detected, 2) in the normal state, and 3)in the deeply overdoped
region where the spin mode is neither expected nor observed.
However this interpretation also requires that the electron-phonon
interaction to be highly anisotropic and its impact on the
electrons to be strongly enhanced in the superconducting state.
This is something one does not expect a priori.

This paper investigates the anisotropy of the electron-phonon
interaction of two known phonons - the half-breathing mode and the
$B_{1g}$ buckling mode - formulated within the same framework as
that used to explain phonon lineshape changes observed via Raman
and neutron measurements\cite{pi_coupling,B1g,breathing,buckling}.
We find that the electron-phonon interactions to be very
anisotropic as a consequence of the following four properties of
the cuprates: 1) the symmetry of the phonon polarizations and
electronic orbitals involved leading to highly anisotropic
electron-phonon coupling matrix elements $g(k,q)$; 2) the
kinematic constraint related to the anisotropy of the electronic
band structure and the van Hove singularity (VHS); 3) the d-wave
superconducting gap, and 4) the near degeneracy of energy scales
of the $B_{1g}$ phonon, the superconducting gap, and the VHS.
Taking into account these and thermal effects we can naturally explain
the observed momentum and temperature dependence of the data,
leading to unified understanding of band renormalizations.

Our starting point is the tight-binding three-band
Hamiltonian modified by
$B_{1g}$ buckling vibrations as considered in Ref.
\cite{buckling2} and in-plane breathing vibrations as considered
in Ref. \cite{Piekarz}:
\begin{eqnarray} \label{Hamiltonian}
&&H=\sum_{\textbf{n},\sigma}\epsilon_{Cu}b^{\dagger}_{\textbf{n},\sigma}b_{\textbf{n},\sigma}\nonumber
\\
&&+\sum_{\textbf{n},\delta,\sigma}\{\epsilon_{O}+eE_{z}u^{O}_{\delta}(\textbf{n})\}
a^{\dagger}_{\textbf{n},\delta,\sigma} a_{\textbf{n},\delta,\sigma}\nonumber\\
&&+\sum_{\textbf{n},\delta,\sigma}P_{\delta}[t-Q_{\delta}g_{dp}u^{O}_{\delta}(\textbf{n})]
(b_{\textbf{n},\sigma}^{\dagger}a_{\textbf{n},\delta,\sigma}
+ h.c.)\nonumber\\
&&+t^{\prime}\sum_{\textbf{n},\delta,\delta^{\prime}}P_{\delta,\delta^{\prime}}
(a_{\textbf{n},\delta}^{\dagger}a_{\textbf{n},\delta^{\prime},\sigma}
+ h.c)
\end{eqnarray}
with Cu-O hopping amplitude $t$, O-O amplitude $t^{\prime}$, and
$\epsilon_{d,p}$ denoting the Cu and O site energies,
respectively. Here $a_{\alpha=x,y},a^{\dagger}_{\alpha=x,y}$
($b,b^{\dagger}$) annihilates, creates an electron in the
O$_{\alpha}$ (Cu) orbital, respectively. The overlap factors
$Q_{\delta}, P_{\delta}, $ and $P_{\delta,\delta^{\prime}}$ are
given in our notation as $Q_{\pm x}=Q_{\pm y}=\pm 1; P_{\pm
x}=-P_{\pm y}=\pm 1; P_{\pm x,\pm y}=1=-P_{\mp x,\pm y}$.
For the half-breathing mode we consider only couplings
$g_{dp}=q_{0}t$ (with $1/q_{0}$ a length scale) arising from
modulation of the covalent hopping amplitude
$t$\cite{Piekarz,footnote}. For the $B_{1g}$ mode a coupling to
linear order in the atomic displacements arises from a local
c-axis oriented crystal field $E_{z}$ which breaks the mirror
plane symmetry of the Cu-O plane\cite{buckling2}.

The symmetry of the $p,d$ wavefunctions of the Cu-O plane is
embedded in the amplitudes $\phi$ of the tight-binding
wavefunctions forming the anti-bonding band with energy dispersion $\epsilon({\bf k})$
arising from Eq.
\ref{Hamiltonian}. Specifically the transformation is determined
by the projected part of the tight-binding wavefunctions
$b_{\textbf{k},\sigma}=\phi_{b}(\textbf{k})d_{\textbf{k},\sigma}$
and
$a_{\alpha,k,\sigma}=\phi_{\alpha}(\textbf{k})d_{\textbf{k},\sigma}$,
\begin{equation}
\label{wavefunction1}
\phi_{b}(\textbf{k})=\frac{1}{N(\textbf{k})}[\epsilon^{2}(\textbf{k})-t^{\prime
2}(\textbf{k})],
\end{equation}
\begin{equation}
\label{wavefunction2}
\phi_{x,y}(\textbf{k})=\frac{\mp
i}{N(\textbf{k})}[\epsilon(\textbf{k})t_{x,y}(\textbf{k})-t^{\prime}(\textbf{k})t_{y,x}(\textbf{k})],
\end{equation}
with  $N^{2}(\textbf{k})=[\epsilon^{2}(\textbf{k})-t^{\prime
2}(\textbf{k})]^{2}+
[\epsilon(\textbf{k})t_{x}(\textbf{k})-t^{\prime
}(\textbf{k})t_{y}(\textbf{k})]^{2}+
[\epsilon(\textbf{k})t_{y}(\textbf{k})-t^{\prime
}(\textbf{k})t_{x}(\textbf{k})]^{2}$,
$t_{\alpha}(\textbf{k})=2t\sin(k_{\alpha}a/2),$ and
$t^{\prime}(\textbf{k})=4t^{\prime}\sin(k_{x}a/2)\sin(k_{y}a/2)$.

The projected wavefunctions act in conjunction with the phonon
eigenvectors to determine the full symmetry of the electron-phonon
coupling:
$H_{el-ph}=\frac{1}{\sqrt{N}}\sum_{\textbf{k},\textbf{q},\sigma,\mu}
g_{\mu}(\textbf{k},\textbf{q})c^{\dagger}_{\textbf{k},\sigma}c_{\textbf{k+q},\sigma}
[a^{\dagger}_{\mu,\textbf{q}}+a_{\mu,-\textbf{q}}]$.
Neglecting the motion of the heavier Cu atoms compared to O in the
phonon eigenvectors the form of the respective couplings can be
compactly written as
\begin{eqnarray}
\label{Eq:coupling_constant1} && g_{B_{1g}}({\bf k,q})= eE_{z}
\sqrt{\frac{\hbar}{4M_{O}M({\bf q})\Omega_{B_{1g}}}}
\\
&&\times\left\{\phi_{x}({\bf k})\phi_{x}({\bf
k^{\prime}})\cos(q_{y}a/2)-
\phi_{y}({\bf k})\phi_{y}({\bf k^{\prime}})\cos(q_{x}a/2)\right\}\nonumber\\
\label{Eq:coupling_constant2}&&g_{br}({\bf
k,q})=g_{dp}\sqrt{\frac{\hbar}{2M_{O}\Omega_{br}}}
\sum_{\alpha=x,y}\\
&&\times\left\{\phi_{b}({\bf
k^{\prime}})\phi_{\alpha}(\textbf{k})\cos(k^{\prime}_{\alpha}a/2)-
\phi_{b}(\textbf{k})\phi_{\alpha}({\bf
k^{\prime}})\cos(k_{\alpha}a/2)\right\} \nonumber,
\end{eqnarray}
with $M({\bf q})=[\cos^{2}(q_{x}a/2)+\cos^{2}(q_{y}a/2)]/2$ and
${\bf k^{\prime}}=\textbf{k-q}$. A simple consequence of symmetry
can be observed at small momentum transfers ${\bf q}$ where
$g_{B_{1g}}({\bf k, q=0})\sim \cos(k_{x}a)-\cos(k_{y}a)$, while
$g_{br}\sim \sin(qa)$ for any \textbf{k}. Yet for large momentum
transfer ${\bf q}=(\pi/a,\pi/a)$, $g_{B_{1g}}$ vanishes for all
${\bf k}$ while $g_{br}$ has its maximum for \textbf{k} located on
the nodal points of the Fermi surface. This $q$-dependence has
been the focus before in context with a $d_{x^{2}-y^{2}}$ pairing
mechanism\cite{phonons,Jepsen}, but the dependence on fermionic
wavevector ${\bf k}$ has usually been overlooked.
It is exactly this fermionic dependence which we feel is
crucially needed to interpret ARPES data.

\begin{figure}[]
\centerline{\epsfxsize=3.0in \epsfysize=2.7in \epsffile{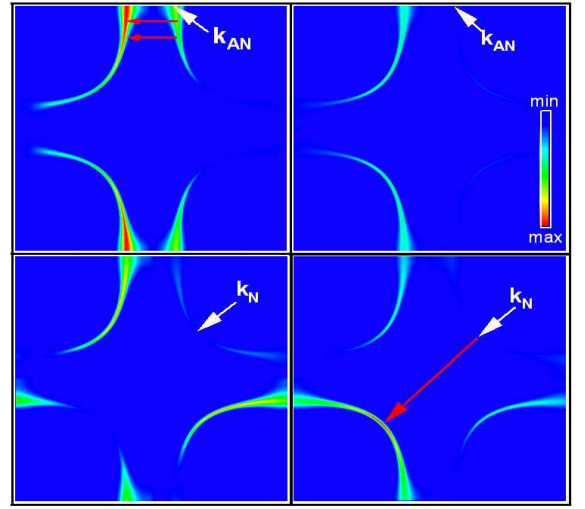}}
\caption{\label{fig: coupling_constant} Plots of the
electron-phonon coupling $\mid g(\textbf{k},\textbf{q})\mid^{2}$
for initial $\textbf{k}$ and scattered $\bf{k^{\prime}=k-q}$
states on the Fermi surface for the buckling mode (left panels)
and breathing mode (right panels) for initial fermion $\textbf{k}$
at an anti-nodal (top panels) and nodal (bottom panels) point on
the Fermi surface, as indicated by the arrows. The red/blue color
indicates the maximum/minimum of the el-ph coupling vertex in the
BZ for each phonon.}
\end{figure}

These features of each coupling constant can be seen in Fig
\ref{fig: coupling_constant} which plots $\mid
g(\textbf{k},\textbf{q})\mid^{2}$ as a function of phonon
scattered momentum $\textbf{q}$ connecting initial $\textbf{k}$
and final $\bf{k^{\prime}=k-q}$ fermion states both on the Fermi surface.
Here we have adopted
a five-parameter fit to the band structure used
previously for optimally doped Bi-2212\cite{neutron}.
For an anti-nodal initial fermion $\textbf{k}_{AN}$, the coupling
to the $B_{1g}$ phonon is largest for $2\textbf{k}_{F}$ scattering
to an anti-nodal final state. The coupling to the
breathing mode is much weaker overall, and is especially weak
for small $\textbf{q}$ and mostly
connects to fermion final states at larger $\textbf{q}$. For
a nodal fermion initial state $\textbf{k}_{N}$, the $B_{1g}$
coupling is suppressed for small $\textbf{q}$ and has a weak
maximum towards anti-nodal final states, while the breathing mode
attains its largest value for scattering to other nodal states,
particularly to state $-\textbf{k}_{N}$.

\begin{figure}[]
\epsfxsize=3.25in \epsffile{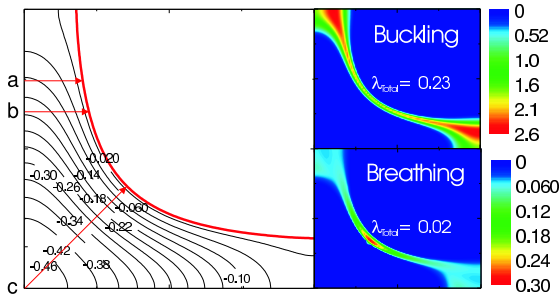} \vskip -6.0cm
\caption{\label{Fig:lambda} Plots of the electron-phonon coupling
$\lambda_{\bf k}$ in the first quadrant of the BZ for the buckling
mode (right top panel) and breathing mode (right bottom panel).
The color scale is shown on the right for each phonon. The left
panel shows energy contours for the band structure
used\cite{neutron}.}
\end{figure}

In Fig. \ref{Fig:lambda} we show effective couplings $\lambda_{\bf k}$
(summed over all ${\bf q}$) for
${\bf k}$ points in the first quadrant of the BZ in
the normal state\cite{Jepsen}.
Here we have used the value of the electric
field $eE_{z}=1.85 eV/\AA$ consistent with Raman scattering
measurements of the $B_{1g}$ phonon and buckling of the Cu-O plane
in YBaCuO, and have taken $q_{0}t=2\sqrt{2}eE_{z}$ as a
representative value for the coupling to the breathing mode. These
values lead to net couplings similar to that obtained from LDA
calculations\cite{Jepsen}. As a result of the anisotropy of the
coupling constant, large enhancements of the coupling near the
anti-node for the $B_{1g}$ phonon, and nodal directions for the
breathing phonon mode are observed, respectively, which are an
order of magnitude larger than the average $\lambda$'s.

We now turn to the spectral function. The dynamic nature of the
coupling is governed by the energy scales of the band structure,
phonon frequencies, and superconducting gap energy, and are
manifest in features of the Nambu-Eliashberg electron-phonon
self-energy\cite{Sandvik}: $\hat\Sigma(\textbf{k},i\omega_{m})=
i\omega_{m}(1-Z(\textbf{k},i\omega_{m}))\hat\tau_{0} +
\chi(\textbf{k},i\omega_{m})\hat\tau_{3}+\Phi(\textbf{k},i\omega_{m})\hat\tau_{1}$
with
\begin{eqnarray}
\label{Eliashberg} &&\hat\Sigma(\textbf{k},i\omega_{m})=
\frac{1}{\beta N}\sum_{\textbf{p},\mu,m}g_{\mu}(\textbf{k,p-k})
g_{\mu}(\textbf{p,k-p})\nonumber\\
&&\times D^{0}_{\mu}(\textbf{k-p},i\omega_{m}-i\omega_{n})
\hat\tau_{3}\hat G^{0}(\textbf{p},i\omega_{m})\hat\tau_{3},
\end{eqnarray}
with $\hat G^{0}(\textbf{p},i\omega_{m})=
\frac{i\omega_{m}\tau_{0}+\epsilon(\textbf{p})\hat\tau_{3}+\Delta(\textbf{p})\hat\tau_{1}}{(i\omega_{m})^{2}-
E^{2}(\textbf{p})}$ and
$E^{2}(\textbf{k})=\epsilon^{2}(\textbf{k})+\Delta^{2}(\textbf{k})$,
with $\hat\tau_{i=0\cdots 3}$ are Pauli matrices. We take
$\Delta(\textbf{k})=\Delta_{0}[\cos(k_{x}a)-\cos(k_{y}a)]/2$ with
$\Delta_{0}=35$meV. The bare phonon propagators are taken as
$D^{0}_{\mu}(\textbf{q},i\Omega_{\nu})=\frac{1}{2}\left[
\frac{1}{i\Omega_{\nu}-\Omega_{\mu}}-
\frac{1}{i\Omega_{\nu}+\Omega_{\mu}}\right].$ Here we take
dispersionless optical phonons $\hbar\Omega_{B_{1g},br}=36,70$meV,
respectively.

The spectral function $A(\textbf{k},\omega)=-\frac{1}{\pi}Im
G_{11}(k,i\omega\rightarrow\omega+i\delta)$ is determined from Eq.
\ref{Eliashberg} with $\hat
G(k,i\omega_{m})=[\hat\tau_{0}i\omega_{m}Z(\textbf{k},i\omega_{m})-(\epsilon(\textbf{k})+
\chi(\textbf{k},i\omega_{m}))\hat\tau_{3}-\Phi(\textbf{k},i\omega_{m})\hat\tau_{1}]^{-1}$
and the coupling constants given by Eq.
\ref{Eq:coupling_constant1}-\ref{Eq:coupling_constant2}. Here we
consider three different cuts: along the nodal direction
($k_{x}=k_{y}$, denoted by \emph{c}), and two cuts parallel to the BZ
face ($k_{y}=0.75\pi/a$, \emph{b})
and ($k_{y}=0.64\pi/a$, \emph{a})
where bilayer effects are not severe\cite{Tanja},
as indicated in Fig. \ref{Fig:lambda}.

\begin{figure*}[]
\centerline{\psfig{file=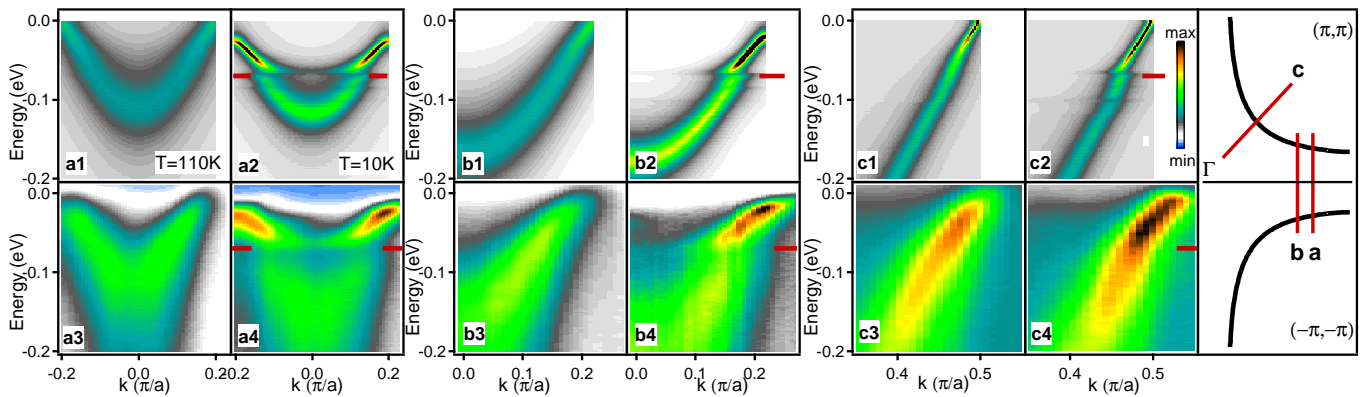,width=18cm,silent=}}
\caption{\label{Fig:3} Image plots of the calculated spectral
functions in the normal (a1,b1,c1) and superconducting (a2,b2,c2)
states compared to the spectral functions in the normal (a3,b3,c3)
and superconducting (a4,b4,c4) states measured in
Bi$_{2}$Sr$_{2}$Ca$_{0.92}$Y$_{0.08}$Cu$_{2}$O$_{8+\delta}$
(Bi-2212)\cite{Tanja} for momentum cuts {\emph{a,b,c}} shown in
the right-most panel and in Figure 2. The same color scale is used
for the normal/superconducting pairs within each cut, but the
scaling for the data and the calculation are separate. The red
markers indicate 70meV in the superconducting state.}
\end{figure*}

Our results for the three cut directions are shown in Fig
\ref{Fig:3} for the normal ($T=110K$) and superconducting
($T=10K$) states.
One can only observe very weak kink features in the normal state,
but a much more pronounced kink at 70 meV in the superconducting
state is observed for all the cuts,
in excellent agreement with the data\cite{Tanja}, shown
also in Fig. \ref{Fig:3}. The dramatic temperature dependence
stems from a substantial change in the electronic occupation distribution
and the opening of the superconducting gap.  In the normal state,
the phonon self energy is a Fermi function at
110K centered at the phonon frequency, which results in a thermal
broadening of 4.4k$_{B}$T or
50meV, comparable to phonon frequencies.
At 10K in the superconducting state, on the other hand, the
phonon self energy is sharply defined due to the
step-function-like Fermi function and a singularity in the
superconducting density of states.  These factors, together with
the experimental difficulty in detecting a weak effect, conspire
to leave the false impression that the bosonic mode merges below
T$_{c}$, as invoked by the spin resonance scenario\cite{neutron}.

In the superconducting state, the data and the theory exhibit a rich variety
of signatures in the electron-phonon coupling: cut $a$ exhibits
the strongest signatures of coupling, showing an Einstein like break
up into a band that follows the phonon dispersion, and one that
follows the electronic one; in cut $b$, a weaker signature of coupling
manifests itself in an $s-$like renormalization of the bare band that
traces the real part of the phonon self-energy; and in cut $c$, where the
coupling to the $B_{1g}$ phonon is weakest, the electron-phonon coupling
manifests itself as a change in the velocity of the band.
The agreement is very good, even for the
large change in the electron-phonon coupling seen between cuts $a$ and $b$,
separated by only $1/10$th of the BZ.

We now discuss the reason for the strong anisotropy as a
result of a concurrence of symmetries, band structure, $d-$wave
energy gap, and energy scales in
the electron-phonon problem. In the normal state, weak kinks are
observed at both phonon frequencies, but more pronounced
at 70 meV for cut \emph{c} and at 36 meV
for cut \emph{a}, as expected from the
anisotropic couplings shown in Fig. \ref{fig: coupling_constant}.
However this is also a consequence of the energy scales of the
phonon modes and the bottom of electronic band along the cut
direction as shown in Fig. \ref{Fig:lambda}. Further towards the anti-node,
another interesting effect occurs.
As one considers cuts
closer to the BZ axis, the bottom of the band along the cut rises
in energy and the breathing phonon mode lies below the bottom of
the band for ($k_{x}=0,k_{y}>0.82\pi/a$) and so the kink feature
for this mode disappears.
This is the usual effect of a Fano redistribution of spectral
weight when a discrete excitation lies either within or outside of
the band continuum\cite{Fano}. This thus can be used to open
"windows" to examine the coupling of electrons to discrete modes
in general.
Taken together,
this naturally explains why the normal state nodal kink is near 70 meV\cite{Nodal}
while the anti-nodal kink is near 40 meV\cite{Anti-Nodal,Tanja}.

In the superconducting state the bare band renormalizes downward
in energy due to the opening of the gap and gives the strongest
effect for the anti-nodal region where the gap is of the order of
the saddle point energy. This also gives the appearance of an increase
of bandwidth.
The $B_{1g}$ coupling becomes dramatically enhanced by the opening of
the gap at a frequency close to the frequency of the phonon
itself, and leads to the dramatic renormalization of the phonon
observed in Raman and neutron measurements\cite{B1g}. Yet the
breathing coupling does not vary dramatically as the large gap
region is not weighted as heavily by the coupling constant. As a
consequence, the $B_{1g}$ mode dominates and
clear kinks in the data are revealed at around 70 meV - the $B_{1g}$ energy
plus the gap.
Indeed a much weaker kink at 105 meV from the
breathing phonon shows at higher energies for the nodal cut but
this kink becomes weaker away from nodal directions as the
coupling and the band bottom along the cut moves below the phonon frequency so
that the ``energy window'' closes.

Our work is formulated in the context of band structure and one phonon
process.  Fig. \ref{Fig:3} indicates that such an approach catches the key features of
the band renormalization effects due to optical phonons in the 35-70 meV
range, something unexpected a priori given the strong correlation effects
in these materials.  Additional broadening in the data of Fig. \ref{Fig:3}
is likely a manifestation of the Coulomb correlation over a broad energy range. A
next step is to incorporate the correlation effect, vertex correction, and
possible non-linear phonon effects, especially as we push towards
underdoped regime.  These are challenges ahead.  On the other hand, the
current approach allows us to study the doping trend in the overdoped
regime.

In summary, contrary to usual opinion of the role of anisotropy in
electron-phonon coupling, the interplay of specific coupling
mechanisms of electrons to the buckling and breathing phonons give
a natural interpretation to the bosonic renormalization effects
seen in ARPES in both the normal and superconducting states, and
provides a framework to understand renormalizations as a function
of doping.


\begin{thebibliography}{99}
\bibitem{footnote}
The coupling $g_{d}$ arising from modulation of Cu site energies
leads to a coupling which only depends on momentum transfers on
not on fermion momentum\cite{Piekarz}.
\end{thebibliography}
\end{document}